# NEW MEASUREMENTS OF THE TRANSITION TO THE NORMAL STATE INDUCED BY HIGH CURRENT DENSITIES IN HIGH-Tc SUPERCONDUCTOR MICROBRIDGES UNDER THERMAL SMALLNESS CONDITIONS.

J.M. Doval[1,a], J. Maza[1,b], C. Torron[1,c], J. A. Veira[1,d], M. Tello[2, e], F.Vidal[1,f]

[1]LBTS. Universidad de Santiago de Compostela, E15782, Spain.

[2]Facultad de Ciencias, Universidad del País Vasco, Bilbao, E48080, Spain.

**Abstract.** We address here the superconductivity quenching under an external magnetic field of amplitudes up to 1 T and in the so-called "thermal smallness" condition, when the microbridge width becomes smaller than the thermal diffusion length of both the own superconductor and its refrigerant (the substrate, in the case of thin films), which breaks their thermal dimensional scaling. Our results further support that when the current perturbations have characteristic times in the millisecond range the quenching is due to thermal instabilities associated with regular (nonsingular) flux-flow, and they also suggest how to optimize the refrigeration of practical superconductors.

[a]juanmanuel.doval@usc.es, [b]jesusj.maza@usc.es, [c]carolina.torron@usc.es, [d]antonio.veira@usc.es, [e]manuel.tello@ehu.es, [f]felix.vidal@usc.es (main corresponding author).

# Introduction

The behavior of the superconductors submitted to electrical current densities well above the critical current density, Jc, where the dissipation first appears, up to current densities (denoted J*) capable of induced a complete transition to the normal state (quenching phenomena), is a long standing but still open problem, in spite of its considerable interest from both the fundamental point of view and for some of the most promising technological applications. These applications concerns, for instance, electrical energy transport under high current densities or superconducting fault current limiters. Nevertheless, even the most general aspects of the superconductivity quenching physics are at present not well settled and under debate. Earlier results on that issue may be seen in Refs. 1 to 9. For more recent results see, e. g., Refs. 10 to 15.

As summarized in Refs. 14 and 15, the quenching mechanisms proposed so far may be crudely classified in two classes: i) Electrodynamic effects, which include the well known flux-flow vortex instability mechanism, first proposed almost forty years ago. In this case the quenching phenomena will be directly driven by the electrical current, and the possible small temperature increases are irrelevant in that even at strictly no overheating the voltage jump would be triggered at J*. ii) The other mechanism invokes heat-driven thermal instabilities, based on the idea that a small increase in temperature due to (nonsingular) magnetic vortex motion triggers a thermal runaway. Note also that the relevance of these different mechanisms, and even their possible entanglement, may depend of the experimental conditions, in particular the characteristic times of the current perturbation.

In contrast with any electrodynamic origin for the quenching, the thermal instabilities will be deeply affected by superconductor refrigeration conditions. The tuning of these thermal conditions will provide, then, a direct way to distinguish between both types of mechanisms. As demonstrate experimentally by Ruibal and coworkers a few years ago [10], this can be done by using superconducting thin film microbridges with widths, $w$, smaller than the thermal diffusion length of both the microbrige and its substrate, $\lambda_{th}$. Under such a condition, that we have coined "*thermal smallness*", the heat exchange deeply depends on the microbrige width, and the conventional thermal dimensional scaling is broken [16]. These measurements, performed with micrrobridges of different widths, but always below 100 μm, and under pulsed electrical currents of 1 millisecond width (which fulfill well the thermal smallness condition, $\lambda_{th}/w > 1$), has provided qualitative but direct experimental evidence of the thermal instability origin of the corresponding quenching [10].

We have now extended the Ruibal and coworkers results [10] by studying the magnetic field dependence of the quenching in $Y_1Ba_2Cu_3O_{7-\delta}$ thin film microbridges, always submitted the current perturbations with characteristic times in the millisecond range and under thermal smallness conditions, but now under external magnetic fields, H, of amplitudes up to 1 T applied perpendicularly to the film ab-planes. Some of these results, which confirm the thermal instability origin of the observed quenching but also open new questions, are now summarized here.

## Experimental details and results

As in the experiments of Ruibal and coworkers [10], the high-$T_c$ superconducting thin films that we have used now were again of the prototypical cuprate superconductor $YBa_2Cu_3O_{7-\delta}$, grown over $SrTiO_3$ substrates. The thermal diffusion length for thermal perturbations with characteristic times in the millisecond range were of the order of 150 μm for the films, and 250 μm for the substrates [10]. This last thermal length is particularly important because it is through the substrate that most of the heat produced in a film is evacuated. As the implemented microbridges had widths between 5 μm and 100 μm and thickness 120 nm, the heat exchange between the films and their substrates was well inside the thermal smallness regime. Another aspect to be stressed here is that these microbridges with different widths have quite similar structural and electrical properties. This last was a crucial requisite to resolve experimentally the possible thermal smallness effects.

In Fig.1, it is presented the dependence on the microbridge width, *w*, of the relationship between the so-called supercritical current density, J*, at which the quenching arise, and the critical current density, Jc, where the dissipation first manifest (originating then a measurable transversal electrical field). These two curves correspond to the average of the data points measured not only in our present work but also those published in Fig.4 of Ref. 11 for microbridges of similar composition and electrical behavior under current densities well below J*. The electrical current pulses used in these measurements have also in both cases a time-width of 1 millisecond. This leads to $\lambda_{th} \approx$ 150 μm and 250 μm for the YBCO microbridges and, respectively, their substrates [10]. The upper curve was measured under an external magnetic field, applied perpendicularly to the films ab-planes, of 1 T, whereas the lower curve was measured in absence of an external magnetic field. The normalized critical temperature, T/Tc, was 0.8 for all data points.

Both curves in Fig. 1 show an increase of J*/Jc when the thermal smallness increases (when *w* decreases and, then, $\lambda_{th}/w$ increases). As already stressed in Ref. 10 when analyzing the results in zero apply magnetic field, such a behavior may be easily understood at a qualitative level on the grounds of the thermal runaway scenario: just taking into account that under smallness conditions the substrate volume involved per unit time in the heat exchange with the superconductor microbridge is almost proportional to the substrate thermal diffusivity length and independent of the microbrige width, *w*. In contrast, the total thermal dissipation, very important near J*, remains in all cases proportional to *w*. So, under thermal smallness conditions the ratio between microbridge refrigeration and dissipation will appreciably improve as w decrease. In contrast with Jc, that will not be appreciably affected, if the quenching is due to thermal instabilities associated with regular flux-flow, J* will appreciably increase as *w* decrease, given rise to the J*/Jc behavior observed in Fig. 1.

By using an analytical approach for the thermal instabilities, a quantitative account of the behavior of J* versus *w* in absence of an external magnetic field was recently proposed in Refs. 13 and 14 on the grounds of the thermal runaway scenarios. However, the magnetic field dependence of J* remains relatively unexplored up to now. As a further contribution to that issue, in Fig. 2 it is presented an example of the magnetic field dependence of both Jc and J*. This example corresponds to a microbridge of 24 μm and for T/Tc = 0.94.

The results of Fig. 2 (a) shows that J* is less affected than Jc by the apply magnetic field, leading then to an important increase of the report J*(H)/Jc(H) when H increases, as showed in Fig. 2(b). Such a behavior is not incompatible with a thermal instability origin of the quenching. In fact, as the temperature dependence of the dissipation per unit volume, W, is a crucial ingredient for a thermal runaway [6, 10, 13, 14], the behavior of J*(H)/Jc(H) could be associated, at least in part, with the softening observed in our experiments, for similar values of W, of dW/dT when applying a magnetic field. The quantitative understanding of the H-dependence of J*, including the possible influence of indirect electrodynamic effects (e.g., inducing current redistributions) on the own thermal instabilities, remains at present a central open question.

# Conclusions

In addition to its interest for the understanding of the superconducting behavior under high current densities and the quenching phenomena, the results summarized here extend to the presence of external magnetic fields the practical rule, already suggested without apply field, to increment the quenching (supercritical) current densities when implementing superconducting cables and wires [16]: Working under "thermal smallness". This could be implemented by using multi-filamentary conductors with diameters of each filament below the thermal diffusion length corresponding to the working conditions. For industrial electrical currents, this practical rule will be compatible with the well-established procedure followed since many years ago to optimize the commercial superconducting wires and to increase also their critical current density: The use of multifilamentary cables with diameters of each filament below the magnetic penetration length. This increases also the surface/volume ratio, increasing then the surface pinning that in turn increases the critical current density. Other practical implications of the thermal smallness conditions to improve the superconducting fault current "microlimiters" may be seen in Ref.17.

# Acknowledgements

This work was supported by the Spanish MICIN under contract ERDF FIS2010-1908.

# Figures

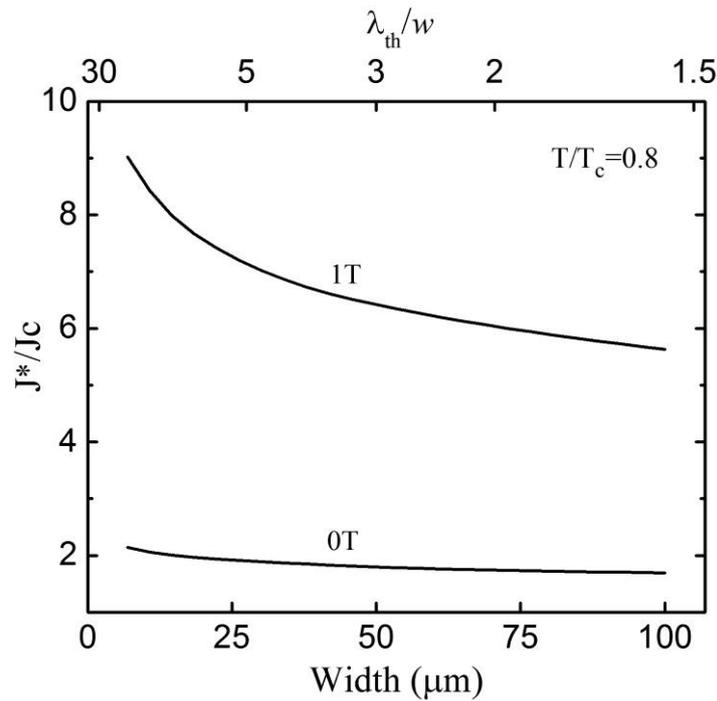

Figure 1. Microbridge width dependence of the quotient between supercritical current density (J*), at which the quenching arise, and the critical current density (Jc). The upper and lower curves correspond to the average data points measured under an external magnetic field of 1T and, respectively, without apply magnetic field. As shown by the upper scale, these measurements are well in the thermal smallness regime.

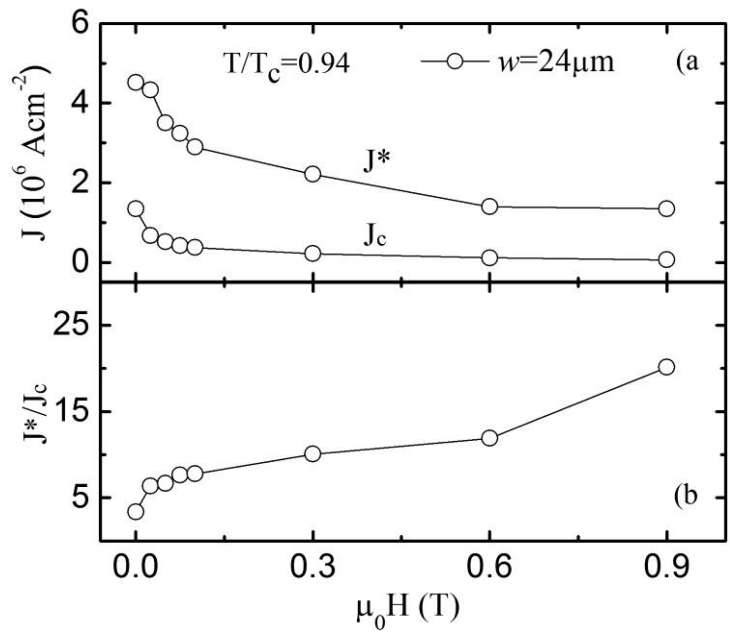

Figure 2. Magnetic field dependence of the critical and the supercritical current densities (a), and of their quotient (b), for a microbridge of 24 μm width under current pulses of 1 millisecond width and for a reduced temperature of 0.94.